\documentclass{article}

\usepackage[preprint]{neurips_2026}

\usepackage[utf8]{inputenc}
\usepackage[T1]{fontenc}
\usepackage{hyperref}
\usepackage{url}
\usepackage{booktabs}
\usepackage{threeparttable}
\usepackage{amsfonts}
\usepackage{amsmath}
\usepackage{amssymb}
\usepackage{nicefrac}
\usepackage{microtype}
\usepackage[table]{xcolor}
\definecolor{libbase}{HTML}{F5F5F5}
\definecolor{liblast}{HTML}{E8F5E9}
\usepackage{graphicx}
\usepackage{subcaption}
\usepackage{algorithm}
\usepackage{listings}
\graphicspath{{figures/}}

\title{Libra: Training the Environment for Agentic Information Retrieval}

\author{%
  Xuan Zhao\\
  \texttt{xuan.zhao@salesforce.com}
  \And
  Andy Chiu\\
  \texttt{andy.chiu@salesforce.com}
  \And
  Gengyu Wang\thanks{Corresponding author.}\\
  \texttt{gengyu.wang@columbia.edu}
}

\begin{document}

\maketitle

\begin{abstract}
Information localization within massive repositories is a cornerstone of agentic LLM systems. While synthetic data-driven optimization has proven successful in training LLMs, little attention has been paid to optimizing the agent's working environment (the repository itself) in a data-driven manner. To bridge this gap, we present \textsc{Libra}, a self-evolving framework that introduces mutable ``catalogs'' (hierarchical Markdown files serving as navigable indices) into the repository. \textsc{Libra} runs an LLM-driven optimization loop where a Prompter generates synthetic queries, a frozen Solver attempts to resolve them by navigating the catalogs, and a Healer rewrites the catalogs in response to the Solver's localization failures. Evaluations across 12 \textsc{SWE-bench Lite} repositories demonstrate that this environmental healing yields continual, logarithmic improvements in code localization accuracy. Furthermore, these environmental improvements transfer zero-shot across different LLMs and problem sets. Although the focus of this paper is to study the general behavior of such a system, we also demonstrate that a minimalist coding agent equipped with \textsc{Libra}-optimized catalogs outperforms state-of-the-art baselines. Code is available at \url{https://github.com/salesforce-misc/Libra} and data at \url{https://huggingface.co/datasets/Salesforce/Libra}.
\end{abstract}

\section{Introduction}
\label{sec:intro}

\begin{figure}[t]
  \centering
  \includegraphics[width=0.95\linewidth]{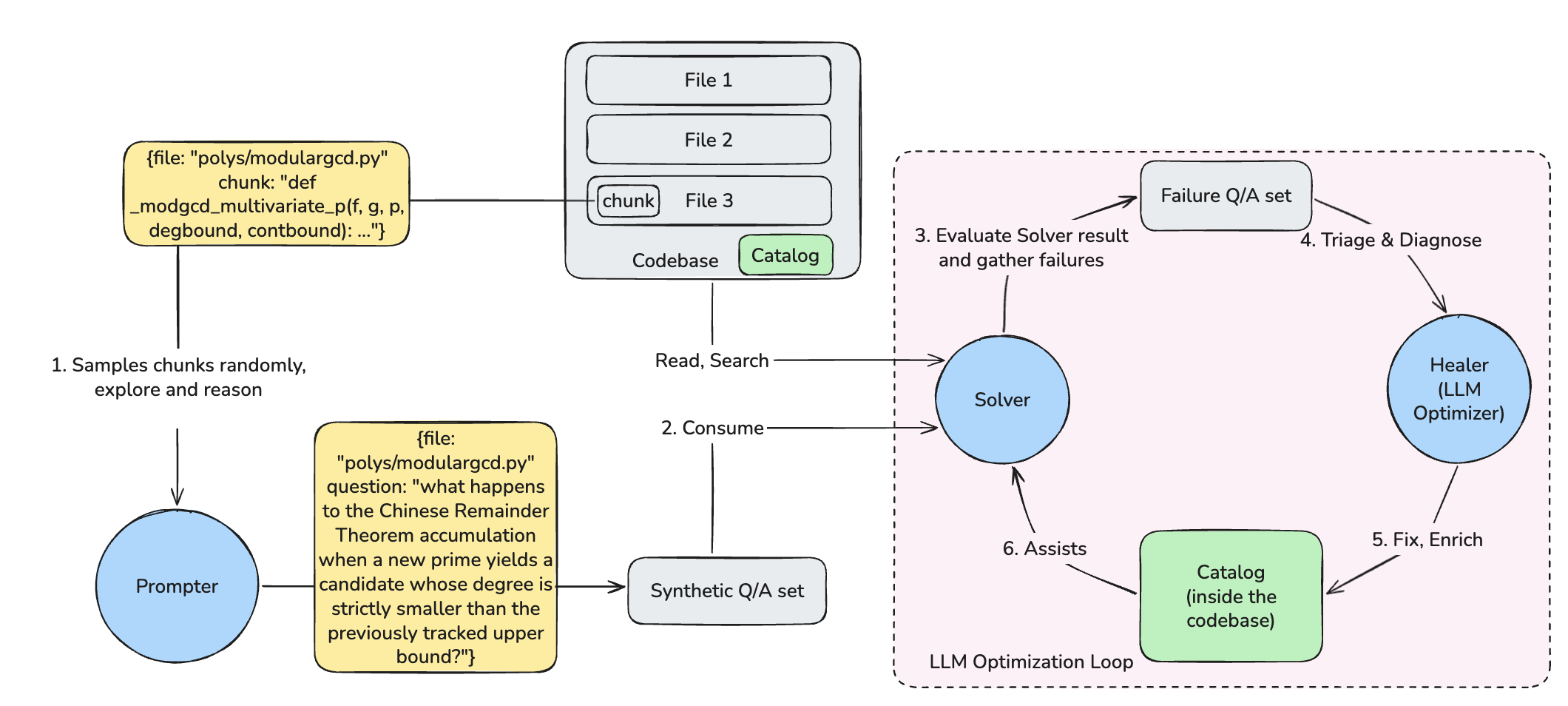}
  \caption{Overview of the Libra system. Three frozen agents
  (Prompter, Solver, Healer) drive an adversarial loop in which only
  the Markdown catalog is updated. The Prompter sees a random file
  chunk and fabricates a question whose answer is that chunk; the
  Solver does not see the chunk and must answer by searching the
  repository alongside the catalog; the Healer reads the accumulated
  failure report once per batch and edits the catalog to repair the
  routing signal.}
\end{figure}

Code localization within large repositories is a fundamental capability for autonomous coding agents~\citep{anthropic2024claude,cursor2024,yang2024sweagent,openhands2024}. Localization accuracy has further been shown to correlate strongly with downstream task resolution rates~\citep{locagent2025,repomem2025}, making it one of the key contributors to end-to-end agent performance. More broadly, information retrieval serves as the cornerstone of agentic systems and has consequently been the subject of extensive research~\citep{lewis2020rag,jin2025searchr1,zhang2026agenticrag}. These agentic systems generally operate on three interacting pillars: the underlying foundation \emph{model}, the \emph{agent design} (prompt engineering and tool use), and the \emph{environment} itself (repository contents, search indices, and documentation). 

To improve an agent's ability to navigate a specific environment, a natural approach is to fine-tune the underlying foundation model using repository-derived data~\citep{jimenez2024swellama,ma2024lingma,pan2024swegym,wei2025swerl}. While this allows the model to internalize environmental knowledge into its weights using mature optimization techniques, it inherently locks the system to a specific model. Consequently, the agent may lose access to the superior reasoning and semantic capabilities of state-of-the-art commercial LLMs, rendering this approach sub-optimal for practical, evolving applications.

Alternatively, researchers have sought to augment the agent's environment by constructing static or reverse indices~\citep{jin2025searchr1,locagent2025,repomem2025,zhang2026agenticrag} or hierarchical summaries~\citep{sarthi2024raptor,edge2024graphrag} and equipping the agent with sophisticated tools to interact with these structures. While these environment-centric approaches are model-agnostic by design, they remain static, lacking a mechanism to adapt dynamically to the agent's specific retrieval challenges or to leverage data---such as human feedback or synthetic interaction traces---to continuously improve the system.

A separate line of work leverages interaction trace data, allowing agents to learn from their own failures through adversarial fine-tuning~\citep{yue2026drzero} or by maintaining persistent memory artifacts within the environment~\citep{he2025sweadept,liu2025experepair}. Despite these advancements, the field still lacks a \emph{model-agnostic}, \emph{data-driven} framework that can \emph{actively} evolve an agent's information retrieval capabilities for a specific repository.

To bridge this gap, we introduce Libra, representing a paradigm shift in repository indexing: moving from static, external embeddings to an actively evolving, plain-text index natively integrated into the repository. Libra constructs a hierarchy of Markdown files, termed \emph{catalogs}, that acts as a navigation map. Instead of relying on predefined heuristics, these catalogs are iteratively optimized through an adversarial LLM-driven training loop. A Prompter agent generates synthetic queries from random code chunks, while a Solver attempts to answer them using only the catalogs for guidance. The Solver's navigation failures provide a rich, targeted training signal for a Healer agent to continuously refine the catalogs, aligning the index structure with how language models actually search for information.\footnote{In our implementation, we execute the Prompter separately to generate the complete training dataset upfront. The Solver then processes these questions in batches, and the Healer analyzes the aggregated failure signals from each batch to update the catalogs.}

We evaluate Libra across the 12 Python repositories of \textsc{SWE-bench Lite}. Our experiments demonstrate the efficacy of this LLM optimization approach: training on synthetic queries yields significant improvements in navigation accuracy that robustly transfer to resolving real-world \textsc{SWE-bench Lite} issues. Furthermore, we show that the fully-trained catalogs act as a persistent, model-agnostic resource, consistently enhancing the localization performance of various downstream models. Compared to state-of-the-art index-based approaches like LocAgent and RepoMem, Libra achieves competitive performance while offering the distinct advantages of interpretability and portability.
\section{Related work}
\label{sec:related}

\paragraph{Code localization and repository indices.}
Repository-scale localization on benchmarks like SWE-bench is dominated
by two families: agentic search over the source tree
(SWE-agent~\citep{yang2024sweagent}, Agentless~\citep{xia2024agentless},
LocAgent~\citep{locagent2025}, RepoMem~\citep{repomem2025}) and
fine-tuned retrievers or models trained on repository
data~\citep{reddy2025swerank,xie2025swefixer,jimenez2024swellama,wei2025swerl,pan2024swegym}.
Hierarchical or graph-structured retrievers
(RAPTOR~\citep{sarthi2024raptor}, GraphRAG~\citep{edge2024graphrag},
HippoRAG~\citep{gutierrez2024hipporag}) impose multi-resolution structure
on the corpus by clustering or LLM extraction.
All of these construct their index, graph, or model weights once and never
repair them from query-time failures; Libra keeps the agentic-search
interface but turns the index into a state variable updated by Solver
failures.

\paragraph{Synthetic Q/A generation.}
LLM-generated queries have a long history as training data for dense
retrievers (Doc2Query~\citep{nogueira2019doc2query},
InPars~\citep{bonifacio2022inpars},
Promptagator~\citep{dai2023promptagator}, GPL~\citep{wang2022gpl}) and
for bootstrapping instruction corpora
(Self-Instruct~\citep{wang2023selfinstruct},
Evol-Instruct~\citep{xu2024wizardlm}). The closest precedent for our
information-asymmetry framing is Beat-the-AI~\citep{bartolo2020beatai},
where annotators exploit oracle access to write hard questions.
Libra's Prompter follows this lineage.

\paragraph{Self-improving agents and persistent memory.}
Many systems improve an agent in place by editing its scratchpad
(Reflexion~\citep{reflexion2023}, Self-Refine~\citep{madaan2023selfrefine},
STaR~\citep{zelikman2022star}), by growing per-agent libraries or memory
stores (Voyager~\citep{voyager2023}, MemGPT~\citep{packer2024memgpt},
A-MEM~\citep{xu2025amem}), or by routing self-generated trajectories back
into model weights (AgentTuning~\citep{zeng2024agenttuning},
Dr.\,Zero~\citep{yue2026drzero}); for software engineering specifically,
ExpeRepair~\citep{liu2025experepair} and SWE-Adept~\citep{he2025sweadept}
attach dual-memory or git-like stores to the issue-resolution agent.
In all of these the memory lives inside the agent and is read-write by
the agent exclusively; Libra externalizes the memory to a plain-text
environment sharable across agents.

\paragraph{LLMs as optimizers of textual artifacts.}
A growing line of work uses one LLM to optimize a textual artifact that
another LLM later consumes: prompts via direct search
(APE~\citep{zhou2023ape}, PromptBreeder~\citep{fernando2023promptbreeder})
or via critique-as-gradient
(ProTeGi~\citep{pryzant2023protegi},
TextGrad~\citep{yuksekgonul2024textgrad}); solutions conditioned on score
history (OPRO~\citep{yang2024opro}); and entire LLM pipelines
(DSPy~\citep{khattab2024dspy}).
Libra's Healer fits this template and applies it to the realm of repository information retrieval.

\section{Method}
\label{sec:method}

The Libra system consists of three independent, frozen LLM agents---Prompter, Solver, and Healer---and an orchestration mechanism that composes them into an LLM optimization loop.
Let $R$ denote a source repository.
The system produces three artifacts:
\begin{enumerate}
    \item \textbf{Synthetic Q/A pairs.} Generated by the Prompter and split into disjoint train and test sets. Training pairs drive the LLM optimization loop; test pairs evaluate the learned catalogs.
    \item \textbf{Failure set $\mathcal{F}$.} The subset of training queries on which the Solver fails to retrieve the correct answer. This intermediate artifact serves as the learning signal for the Healer.
    \item \textbf{Libra catalogs $K$.} A set of mutable, plain-text Markdown files that encode repository routing information. Once optimized, these catalogs serve as a persistent index to assist any coding agent in navigating $R$.
\end{enumerate}

In this paper, we evaluate the system on fixed repository snapshots. While the plain-text format of the catalogs naturally supports incremental updates as the repository evolves, we leave this extension to future work.

\subsection{Libra catalogs}
\label{sec:method:tree}

The Libra catalogs are a set of Markdown files placed at the repository root and in each submodule directory. Our hypothesis---validated empirically in Section~\ref{sec:experiments}---is that through iterative refinement, an LLM optimizer can encode repository routing information into these files, and that this information transfers to other coding agents, improving their localization performance. The Libra catalogs are human-readable, human-editable, and typically range from 100--1000\,KB when fully populated.

\subsection{Three frozen agents}
\label{sec:method:agents}

We now define the three agents that constitute the Libra system. All three are \emph{frozen}: their system prompts, tool configurations, and underlying model weights remain fixed across all rounds. Only the catalogs $K$ are mutable.

All three agents are tool-using LLM agents that operate over the source repository $R$ as a shared environment. Following standard practice, we write $R$ as an input to each agent's signature to denote tool-mediated read access; the role-specific tool sets (and, for the Healer, additional write access to $K$) are detailed in Appendix~\ref{app:prompts}.

\paragraph{Prompter (Data Synthesizer).}
The Prompter generates synthetic question--answer pairs that probe the Solver's retrieval capability. Let $\mathcal{C}(R) = \{c_1, \ldots, c_N\}$ denote the set of code chunks in repository $R$, and let $\pi(c)$ denote the unique file path containing chunk $c$. Formally,
\[\textsc{Prompter}:\; (c, R) \;\mapsto\; (q, p^*),\]
where $c \sim \text{Uniform}(\mathcal{C}(R))$ is a uniformly sampled chunk, $q$ is a natural-language question simulating a realistic user query answerable from $c$, and $p^* = \pi(c)$ is the gold file path.

The Prompter has full access to the sampled chunk $c$ and may further explore $R$ via its tools to gather surrounding context. Because the question is crafted \emph{with} the answer in hand, correctness of the gold label is guaranteed by construction.\footnote{Strictly, correctness holds to the extent that the Prompter LLM generates a question that is genuinely answerable from the given chunk and not from other locations in the repository. In practice, we observe this assumption to be reliable enough for the precision of this research.}

\paragraph{Solver (Query Resolver).}
The Solver is the coding agent responsible for resolving queries by navigating the repository. It represents the downstream consumer of the catalogs. Formally,
\[\textsc{Solver}:\; (q, K, R) \;\mapsto\; \hat{p},\]
where $q$ is a natural-language query, $K$ is the current catalog state, and $\hat{p}$ is the predicted file path.

While the Solver can freely traverse $R$ via its tools, it has \emph{no access} to the gold chunk or file path; it must locate the answer solely through repository exploration guided by the catalogs. A hypothesis which we later validate empirically is that this asymmetry in information and difficulty in task creates a gap in performance between the Prompter and the Solver, from which the Healer can harvest reliable signals to refine the catalogs.

\paragraph{Healer (LLM Optimizer).}
The Healer refines the catalogs based on Solver failures, functioning as the optimizer in the Libra system. Formally,
\[\textsc{Healer}:\; (\mathcal{F}, K, R) \;\mapsto\; K',\]
where $\mathcal{F} = \{(q_i, p^*_i, \hat{p}_i)\}_{i=1}^{|\mathcal{F}|}$ is the failure set---each entry contains the query $q_i$, the gold path $p^*_i$, and the Solver's predicted path $\hat{p}_i$---and $K'$ is the updated catalog state. Beyond read access to $R$, the Healer also holds write access to $K$, which it edits in place.

\subsection{The LLM optimization loop}
\label{sec:method:loop}

The LLM optimization loop (also referred to as the training loop) orchestrates the three agents over $T$ rounds. Each round consists of a \emph{testing phase}, in which the Prompter and Solver generate a batch of queries and expose failures, followed by a \emph{Healing phase}, in which the Healer updates the catalogs. The procedure is formalized in Algorithm~\ref{alg:libra}.

\begin{algorithm}[t]
\caption{Libra LLM optimization loop.}
\label{alg:libra}
\begin{tabbing}
\quad\=\quad\=\quad\=\quad\=\kill
\textbf{Input:} repository $R$, batch size $B$, number of rounds $T$. \\
\textbf{State:} Catalog set $K$ (directory of Markdown files), initialized empty. \\[4pt]
\textbf{for} $t = 1, \ldots, T$ \textbf{do} \\
\> \emph{// Testing phase} \\
\> $\mathcal{F}_t \leftarrow \emptyset$ \\
\> \textbf{for} $i = 1, \ldots, B$ \textbf{do} \\
\>\> $c_i \sim \text{Uniform}(\mathcal{C}(R))$ \quad
     // sample a code chunk \\
\>\> $(q_i, p^*_i) \leftarrow \textsc{Prompter}(c_i, R)$ \quad
     // generate question \\
\>\> $\hat{p}_i \leftarrow \textsc{Solver}(q_i, K, R)$ \quad
     // resolve query using catalogs \\
\>\> \textbf{if} $\hat{p}_i \ne p^*_i$ \textbf{then}
     $\mathcal{F}_t \leftarrow \mathcal{F}_t \cup \{(q_i, p^*_i, \hat{p}_i)\}$ \\[4pt]
\> \emph{// Healing phase} \\
\> $K \leftarrow \textsc{Healer}(\mathcal{F}_t, K, R)$ \\[4pt]
\textbf{return} $K$
\end{tabbing}
\end{algorithm}
\section{Experiments}
\label{sec:experiments}

We evaluate Libra along five axes. First, we run an extended training trajectory on \texttt{sympy} to establish our core result: catalog training continuously improves code localization accuracy, yielding logarithmic returns over time (\S\ref{sec:exp:main}). We select \texttt{sympy} as our primary testbed because it is the largest repository in the \textsc{SWE-bench Lite} dataset, and its well-structured, evenly distributed submodules make it ideal for evaluating hierarchical catalogs. 
Second, we verify that this improvement pattern holds across all 12 \textsc{SWE-bench Lite} repositories (\S\ref{sec:exp:multi-repo}), which vary significantly in size and structure. Third, we test whether catalogs trained on synthetic Prompter data transfer to real-world problem distributions by replaying on a subset of \textsc{SWE-bench Lite} (\S\ref{sec:exp:cross-problem}).\textsuperscript{\ref{fn:swe-lite-selection}} Fourth, we assess the model-agnosticism of the learned catalogs by replaying the training trajectory with three different Solver setups (\S\ref{sec:exp:cross-model}). Finally, we demonstrate how individual test queries transition from failure to success as the catalog evolves (\S\ref{sec:exp:case}).

\subsection{Setup}
\label{sec:exp:setup}

\paragraph{Data.}
For catalog training, we use synthetic data generated offline by the Prompter. We evaluate in-distribution learning on a held-out test split of this synthetic data. To assess cross-problem generalizability, we additionally evaluate on a subset of the real \textsc{SWE-bench Lite} problem set. Because the original \textsc{SWE-bench Lite} instances (300 in total) each assume a different base commit, we filter the set down to 199 instances that can be evaluated on a single base commit per repository.\footnote{\label{fn:swe-lite-selection}Gold files and functions are derived from each instance's ground-truth patch and validated against the base commit's file tree. We exclude instances where (1)~the referenced file or function does not exist at the base commit, (2)~the referenced file or module has been moved or renamed between the instance's original commit and the base commit, or (3)~the instance requires multi-hop patching spanning multiple files.} The exact number of training, testing, and \textsc{SWE-bench Lite} instances chosen for each repository, along with the base commit used to rebase the \textsc{SWE-bench Lite} instances, is detailed in Appendix~\ref{app:cost} (Table~\ref{tab:cost}). Both the synthetic Prompter splits and the re-anchored \textsc{SWE-bench Lite} subset are released; see Appendix~\ref{app:datasets} for schema and license details.

\suppressfloats[t]
\begin{figure}[t]
  \centering
  \includegraphics[width=0.95\linewidth]{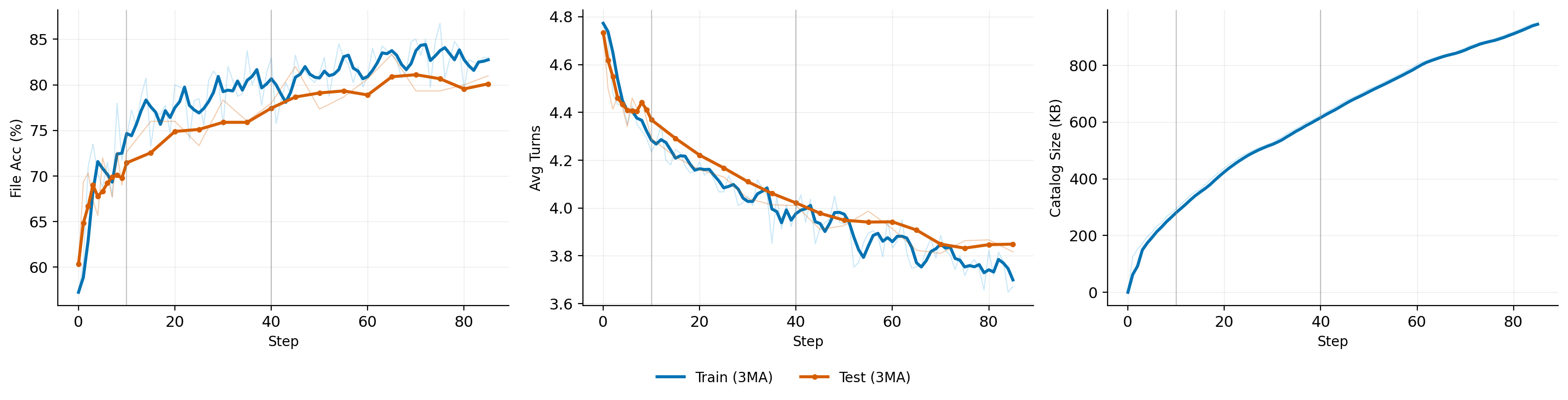}
  \caption{Training trajectory on \texttt{sympy} (85 healer steps, GPT-5-mini at 5 turns). \emph{Left}: file-level accuracy on train and test sets. \emph{Middle}: average turns per query. \emph{Right}: catalog size growth.}
  \label{fig:training-curves}
\end{figure}

\begin{table}[t]
  \centering
  \small
  \caption{Accuracy on the \texttt{sympy} held-out test set (300 instances, GPT-5-mini, 5 turns). LocAgent and RepoMem report mean $\pm$ SEM over 5 runs.}
  \label{tab:accuracy}
  \begin{tabular}{lcc ccc}
    \toprule
    & \multicolumn{2}{c}{Baselines} & \multicolumn{3}{c}{Libra} \\
    \cmidrule(lr){2-3} \cmidrule(lr){4-6}
    Metric & LocAgent & RepoMem & Step 0 & Step 85 & Best (step 65) \\
    \midrule
    File ACC (\%)   & $65.5 \pm 0.7$ & $63.0 \pm 0.4$ & $60.3$ & $81.0$ & $\mathbf{83.3}$ \\
    Func ACC (\%)   & $51.5 \pm 1.1$ & $48.3 \pm 0.6$ & $46.3$ & $70.7$ & $\mathbf{70.7}$ \\
    Cost (\$)\textsuperscript{$\dagger$} & $1.84 \pm 0.02$ & $1.97 \pm 0.02$ & $1.67$ & $4.81$ & $3.83$ \\
    \bottomrule
    \multicolumn{6}{l}{\footnotesize\textsuperscript{$\dagger$}Sum over all 300 instances.}
  \end{tabular}
\end{table}

\paragraph{Agents and Catalog.}
The Solver is powered by GPT-5-mini and operates under a strict budget of \texttt{max\_turns}=5 during training. We constrain the Solver to ensure that routing failures are plentiful, thereby providing the Healer with adequate learning signals (see Appendix~\ref{app:saturation} for a saturation analysis on frontier-tier Solvers that motivates this choice). By design, the Solver follows a deliberately minimalist recipe: it is equipped with only two general-purpose tools---\texttt{Bash} (restricted to \texttt{ls}, \texttt{grep}, and \texttt{find}) and \texttt{Read} (file read with line-range selection)---alongside a short system prompt that instructs it to consult the Libra catalogs. We intentionally avoid bespoke retrieval tooling, repository graphs, or commit-history indices used by prior work (cf.\ the 3-tool LocAgent and 7-tool RepoMem setups in the Baselines paragraph), so that any performance gain can be attributed to the catalogs themselves rather than to richer agent scaffolding. Initially, empty catalogs are created under the repository root and each Python submodule. The root catalog is always provided as context. For our cross-model generalizability evaluation, we maintain this minimalist 2-tool design but swap the underlying model to GPT-5, Gemini-2.5-Flash, or GPT-5-mini with an expanded budget of \texttt{max\_turns}=12.

The Prompter and Healer both utilize Claude Opus~4.6. They are granted full access to the ClaudeAgentSDK tool suite\footnote{Across all training runs, the active tools observed in the agents' trajectories are \texttt{Bash}, \texttt{Read}, \texttt{Write}, \texttt{Edit}, \texttt{Glob}, \texttt{Grep}, \texttt{Agent} (subagent dispatch), and \texttt{StructuredOutput} (final-answer capture). See the Claude Agent SDK documentation at \url{https://docs.claude.com/en/agent-sdk/overview}.} and operate without any turn budget.\footnote{\label{fn:sympy-prompter}The Prompter used to generate the \texttt{sympy} training set is a slightly earlier variant: it also runs Claude Opus~4.6 but is equipped only with \texttt{Bash} and \texttt{Read} tools rather than the full ClaudeAgentSDK.} 

A small set of soft heuristics, all communicated via system prompts rather than programmatically enforced, shape agent 
behavior. The Healer is given structural guidelines that keep catalogs terse and discriminative (a table-like, hierarchical layout with bounded bullet counts and lengths), and the Prompter is instructed to produce harder queries by avoiding exact identifiers and favoring questions that plausibly match sibling chunks. A detailed prompt-engineering study is outside the scope of this paper; the verbatim prompts are reproduced in Appendix~\ref{app:prompts}.

Catalogs are initialized empty for this study. Their content and structural conventions are indirectly shaped by the Healer design.

\paragraph{Baselines.}
We compare against two agentic localization methods: LocAgent~\citep{locagent2025} and RepoMem~\citep{repomem2025}. Like Libra, both are agentic Solvers that explore the repository through tool calls; unlike Libra, neither rewrites the environment as it runs. LocAgent replaces our \texttt{Bash}/\texttt{Read} tools with three structure-aware tools backed by a heterogeneous repository graph (directories, files, classes, functions, and their contain/import/invoke/inherit edges) plus a sparse, hierarchical entity index that falls back from exact-ID/exact-name lookups to BM25~\citep{robertson2009bm25} over entity IDs and code chunks. RepoMem extends LocAgent with four \emph{frozen} commit-history tools that surface episodic and semantic memory mined offline from prior commits and edit frequencies. Both baselines use the same Solver model (GPT-5-mini) and turn budgets as Libra to ensure a controlled comparison; full implementation details are deferred to Appendix~\ref{app:baselines}. All baseline results are averaged over multiple independent runs (5 runs for the \texttt{sympy} test set, 10 runs for the \textsc{SWE-bench Lite} evaluation), and we report the mean $\pm$ SEM.

\paragraph{Metrics.}
We measure accuracy by the exact match between the Solver's prediction and the ground-truth file (\texttt{File ACC}) or function (\texttt{Func ACC}). Because the Solver is constrained to output exactly one prediction, these metrics correspond to the Top-1 accuracy (ACC@1) used in prior work~\citep{locagent2025,reddy2025swerank}. As file and function accuracies are highly correlated (see Appendix~\ref{app:file-vs-func}, Figure~\ref{fig:file-vs-func}), we report file accuracy as our primary metric.

\paragraph{Replay methodology.}
Each Healer step commits the updated catalog to version control, producing a sequence of snapshots that we term the \emph{training trajectory}. To evaluate performance at different stages of training, we \emph{replay} the held-out test set against these catalog snapshots. This approach decouples evaluation from the training loop, allowing us to assess any combination of Solver model, setup, and problem set against any historical catalog state without retraining.

\subsection{Main result: extended training on \texttt{sympy}}
\label{sec:exp:main}

We run an 85-step training trajectory on the \texttt{sympy} repository (the largest repository in \textsc{SWE-bench Lite}, with ${\sim}430$k lines of Python code). Each step processes a batch of 400 training instances, totaling 34{,}000 instances over the full run. The held-out test set contains 300 instances at the same base commit. We also evaluate the baselines LocAgent~\citep{locagent2025} and RepoMem~\citep{repomem2025} on the same test set using the same model and turn budget. 

Table~\ref{tab:accuracy} and Figure~\ref{fig:training-curves} summarize the end-to-end results and the full training trajectory. We observe several key dynamics. First, while the Libra Solver initially underperforms the baselines due to its lack of sophisticated tools and turn-budget efficiency, its performance quickly catches up as the catalogs build up, ultimately surpassing both baselines by a significant margin. Second, the trajectory exhibits \textbf{rapid early gains}: the first 5 steps (2{,}000 instances) lift test file accuracy from $60.3\%$ to $72.0\%$ ($+11.7$\,pp), capturing roughly half of the total improvement. Third, we observe \textbf{logarithmic gains thereafter}, with accuracy climbing in $1$--$3$\,pp increments from step 5 onward and plateauing near step 60 ($80.7\%$). Finally, \textbf{train and test track each other} closely; the gap between train and test accuracy stays within ${\sim}2$\,pp throughout. This confirms our hypothesis that the routing information encoded in the catalog is transferable to problems unseen during training.\footnote{Because we train for only one epoch, every training batch is itself new to the catalog, so even training accuracy measures generalization. We nonetheless maintain a fixed held-out test set for additional clarity.}

\subsection{Training on all \textsc{SWE-bench Lite} repositories}
\label{sec:exp:multi-repo}

To assess Libra's effectiveness across a wider spectrum of repositories varying in size, quality, and structure, we train Libra independently on all 12 repositories selected for \textsc{SWE-bench Lite}. Each repository uses the same configuration (GPT-5-mini, 5 turns), with batch sizes and step counts adapted to the repository's size (see Table~\ref{tab:cost} in Appendix~\ref{app:cost} for exact parameters).

\begin{figure}[t]
  \centering
  \includegraphics[width=0.75\linewidth]{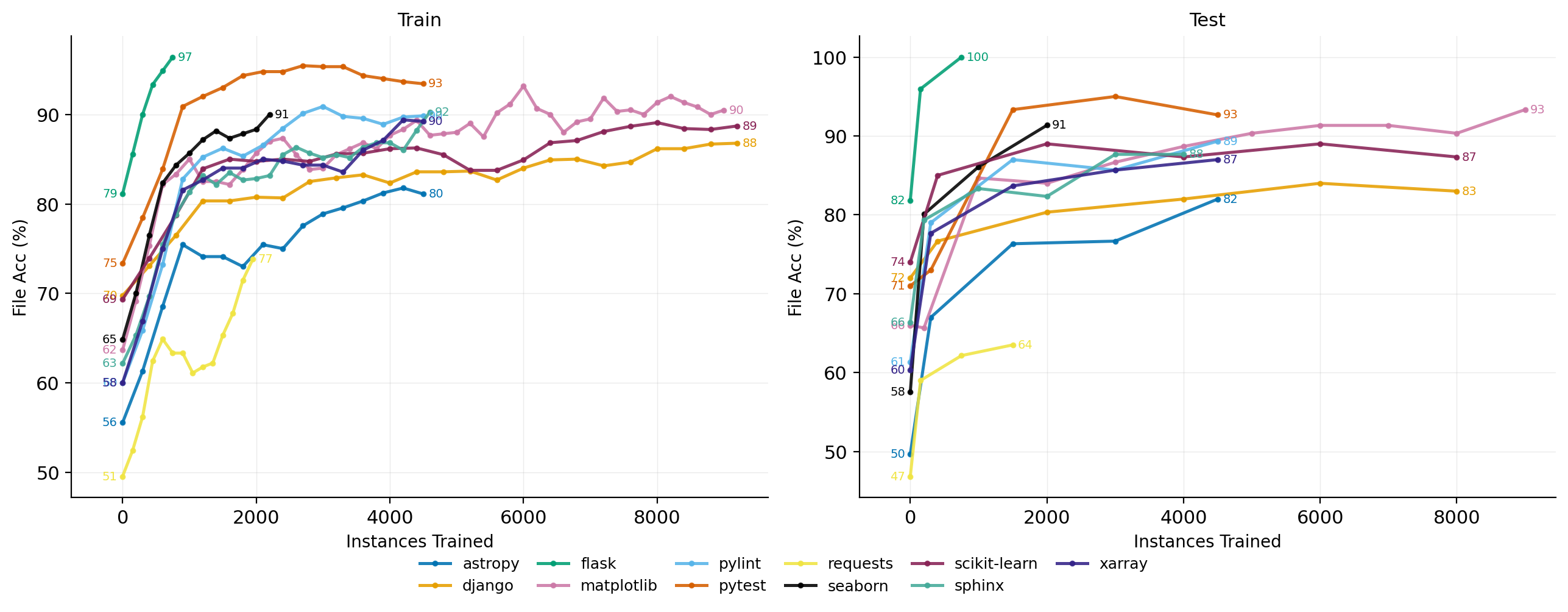}
  \caption{Training curves for each \textsc{SWE-bench Lite} repository (\emph{left}: train, \emph{right}: test).}
  \label{fig:all-repos-curves}
\end{figure}

\begin{table*}[t]
  \centering
  \caption{\textsc{SWE-bench Lite} evaluation (199 instances, GPT-5-mini, 10 runs). Values are mean\,(SEM). The Catalog column indicates Libra's catalog state. Aggregate columns report instance-weighted accuracy and cost across all 12 repos. Per-repo columns report File ACC for the six largest repositories; ``other'' aggregates astropy\,(3), flask\,(3), pylint\,(3), requests\,(2), seaborn\,(4), xarray\,(5).}
  \label{tab:swe-lite}
  \resizebox{\textwidth}{!}{%
  \begin{tabular}{llc ccc @{\hskip 1.5em} ccccccc}
    \toprule
    & & & \multicolumn{3}{c}{Aggregate (199 inst.)} & \multicolumn{7}{c}{Per-repo file accuracy (\%)} \\
    \cmidrule(lr){4-6} \cmidrule(lr){7-13}
    Turns & Agent & Catalog & File ACC (\%) & Func ACC (\%) & Cost (\$)\textsuperscript{$\dagger$} & django\,(78) & sympy\,(52) & mpl\,(15) & pytest\,(13) & sklearn\,(11) & sphinx\,(10) & other\,(20) \\
    \midrule
    \rowcolor{libbase}
    $5$  & Libra              & empty   & $65.8\,(0.8)$ & $53.3\,(0.7)$ & $0.0046$ & $66.2\,(1.6)$ & $56.5\,(1.5)$ & $66.7\,(3.3)$ & $72.3\,(2.6)$ & $80.9\,(2.1)$ & $82.0\,(3.3)$ & $67.0\,(2.1)$ \\
    \rowcolor{libbase}
         & Libra              & bootstrapped    & $73.7\,(0.6)$ & $62.0\,(0.7)$ & $0.0052$ & $74.4\,(0.9)$ & $65.6\,(1.1)$ & $68.0\,(2.2)$ & $75.4\,(3.8)$ & $96.4\,(1.5)$ & $81.0\,(4.1)$ & $79.0\,(1.9)$ \\
         & LocAgent           &         & $76.0\,(0.5)$ & $62.2\,(0.7)$ & $0.0057$ & $\mathbf{79.2}\,(1.1)$ & $65.8\,(1.0)$ & $\mathbf{73.3}\,(2.2)$ & $74.6\,(2.3)$ & $91.8\,(1.6)$ & $72.0\,(2.0)$ & $\mathbf{86.0}\,(1.0)$ \\
    \rowcolor{liblast}
         & \textbf{Libra}     & trained & $\mathbf{77.8}\,(0.5)$ & $\mathbf{66.4}\,(0.5)$ & $0.0102$ & $77.9\,(0.9)$ & $\mathbf{69.2}\,(0.8)$ & $\mathbf{73.3}\,(1.0)$ & $\mathbf{76.2}\,(2.4)$ & $\mathbf{100.0}\,(0.0)$ & $\mathbf{97.0}\,(1.5)$ & $82.0\,(1.9)$ \\
    \midrule
    \rowcolor{libbase}
    $10$ & Libra              & empty   & $77.0\,(0.6)$ & $68.7\,(0.6)$ & $0.0108$ & $77.3\,(1.2)$ & $66.5\,(1.0)$ & $71.3\,(2.2)$ & $83.8\,(1.4)$ & $98.2\,(1.2)$ & $96.0\,(2.2)$ & $82.0\,(1.9)$ \\
    \rowcolor{libbase}
         & Libra              & bootstrapped    & $79.8\,(0.5)$ & $70.5\,(0.5)$ & $0.0109$ & $80.6\,(0.9)$ & $68.3\,(0.9)$ & $75.3\,(2.4)$ & $\mathbf{86.9}\,(2.0)$ & $\mathbf{100.0}\,(0.0)$ & $\mathbf{100.0}\,(0.0)$ & $84.5\,(1.5)$ \\
         & LocAgent           &         & $79.7\,(0.4)$ & $67.7\,(0.4)$ & $0.0113$ & $78.8\,(0.8)$ & $70.4\,(0.7)$ & $76.0\,(2.5)$ & $84.6\,(2.0)$ & $\mathbf{100.0}\,(0.0)$ & $91.0\,(2.8)$ & $90.0\,(1.2)$ \\
         & RepoMem            &         & $79.7\,(0.5)$ & $67.3\,(0.7)$ & $0.0111$ & $79.1\,(0.8)$ & $69.0\,(1.1)$ & $\mathbf{82.0}\,(1.4)$ & $79.2\,(1.6)$ & $98.2\,(1.2)$ & $89.0\,(3.1)$ & $\mathbf{93.5}\,(0.9)$ \\
    \rowcolor{liblast}
         & \textbf{Libra}     & trained & $\mathbf{82.3}\,(0.6)$ & $\mathbf{72.1}\,(0.5)$ & $0.0162$ & $\mathbf{82.3}\,(1.2)$ & $\mathbf{73.5}\,(0.6)$ & $79.3\,(2.5)$ & $85.4\,(2.4)$ & $\mathbf{100.0}\,(0.0)$ & $99.0\,(1.0)$ & $87.0\,(1.4)$ \\
    \bottomrule
    \multicolumn{13}{l}{\footnotesize\textsuperscript{$\dagger$}Average cost per instance.}
  \end{tabular}%
  }
\end{table*}

Figure~\ref{fig:all-repos-curves} confirms that every repository consistently benefits from training. The training curves exhibit the same logarithmic-gains pattern observed on \texttt{sympy}: the first few hundred instances capture the steepest improvement as the Healer bootstraps the empty catalogs. As training progresses, the gains continue and eventually plateau. For some repositories, we observe a slight regression during the final steps, suggesting that the catalogs begin to "overfit" to the training data. This likely occurs because chunk overlaps cause similar questions to appear repeatedly despite single-epoch training, and variable-length catalogs accumulate noise alongside useful information as training progresses.
\subsection{Replay on \textsc{SWE-bench Lite} problem set}
\label{sec:exp:cross-problem}

To verify that catalogs trained on synthetic data improve the Solver's performance on real bug-localization queries, we evaluate the Solver across three catalog healing stages using the \textsc{SWE-bench Lite} problem set: Empty (step 0), Bootstrapped (step 1), and Trained (final step).
We compare these results against our baselines and replicate all experiments with an extended turn budget of 10 to analyze the impact of turn limits across methods. Note that RepoMem is evaluated exclusively at 10 turns, as the LLM requires more than 5 turns to effectively utilize its 7 tools. To account for the limited number of instances in the \textsc{SWE-bench Lite} dataset, we average all measurements across 10 independent runs and report the mean $\pm$ SEM.

Table~\ref{tab:swe-lite} shows consistent gains on real \textsc{SWE-bench Lite} instances as the catalogs are bootstrapped and fully trained. Furthermore, once the catalogs are fully trained, the Libra Solver achieves the best performance among all methods, surpassing both LocAgent and RepoMem in most repositories despite being trained entirely on synthetic data. 

\subsection{Replay on different Solver setup}
\label{sec:exp:cross-model}

A central claim of our work is that training the environment is \emph{orthogonal} to the choice of Solver model and design. To test this, we take the first 60 steps of the catalog trajectory produced by GPT-5-mini training and replay the test set with three different Solver setups: GPT-5-mini (the training Solver) with extended 12-turn budget, GPT-5 (a stronger sibling), and Gemini-2.5-Flash (a different model family). The catalogs are \emph{never re-trained} for the new Solvers. Performance of the original trained Solver is also included as reference.

\begin{figure}[t]
  \centering
  \includegraphics[width=0.35\linewidth]{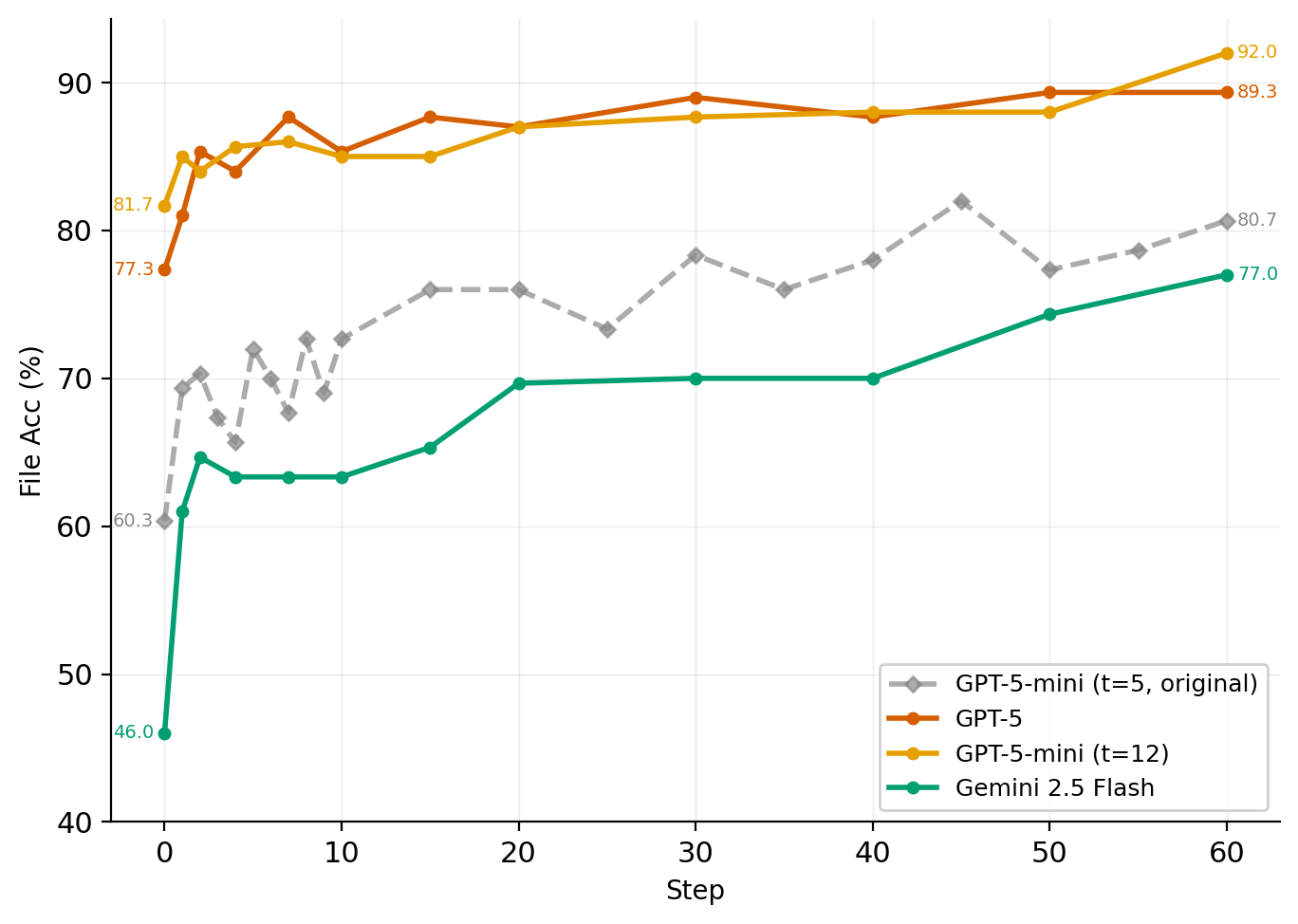}
  \caption{Replay results with different Solver models and turn budgets}
  \label{fig:replay}
\end{figure}

All three Solvers improve along the catalog trajectory (Figure~\ref{fig:replay}). The weaker the no-catalog baseline, the larger the absolute lift: Gemini-2.5-Flash gains $+31.0$\,pp (from $46.0\%$ to $77.0\%$), GPT-5 gains $+12.0$\,pp (from $77.3\%$ to $89.3\%$), and GPT-5-mini gains $+10.3$\,pp (from $81.7\%$ to $92.0\%$). This supports the claim that the catalog is a \emph{model-agnostic} routing artifact whose value compounds with improvements in model capabilities and model cost efficiency.

\subsection{Healing effect on individual instances}
\label{sec:exp:case}

To isolate the steady-state gain, we trace each of the 300 test instances across all evaluation checkpoints and compute two windowed pass rates: an \emph{early} window (steps 0--10, 11 evaluations) and a \emph{later} window (steps 40--85, 10 evaluations).

\begin{figure}[t]
  \centering
  \includegraphics[width=\linewidth]{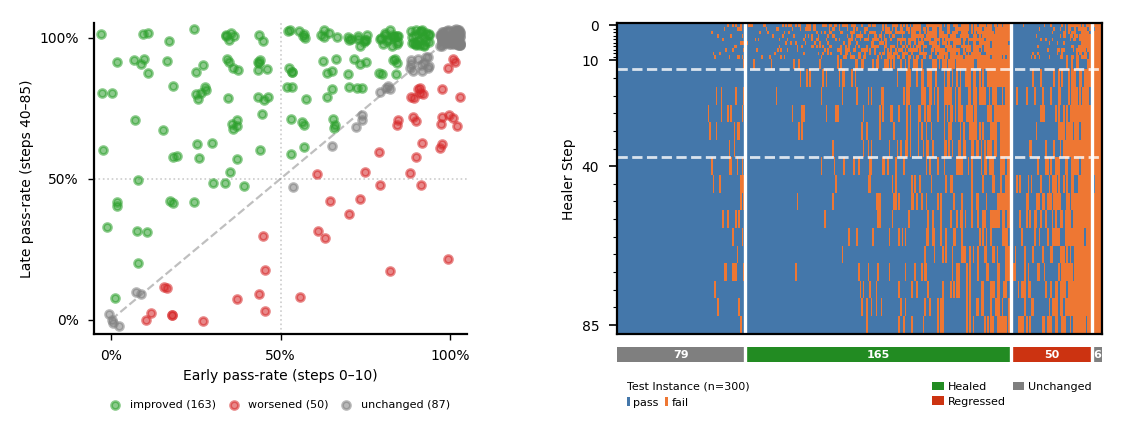}
  \caption{Per-instance analysis of 300 held-out test instances. \emph{Left}: Scatter plot of early pass-rate (steps 0--10) vs.\ later pass-rate (steps 40--85). Instances above the diagonal show improvement, with points further to the top-left indicating stronger healing effects. \emph{Right}: Pass/fail status of each instance throughout the training process. Dashed lines mark the early and later window boundaries. The dominant orange$\to$blue transition in the \texttt{Healed} band is the visual signature of catalog-driven improvement.}
  \label{fig:case-study}
\end{figure}

Figure~\ref{fig:case-study} (left) shows the per-instance view: $163$ of $300$ instances ($54.3\%$) improve from the early to later windows, while only $50$ ($16.7\%$) worsen, yielding a $3.3{:}1$ ratio of improved to degraded performance. The heatmap (Figure~\ref{fig:case-study}, right) visualizes the pass/fail trajectory of each individual instance over the course of training. It highlights a subset of instances (\texttt{healed}) that transition to a consistent pass rate ($>66.7\%$) in the later window, demonstrating the system's ability to systematically correct its routing failures.

\section{Conclusion}
\label{sec:conclusion}

We introduced Libra, a framework that shifts the paradigm of agentic information retrieval from static environment augmentation and model-specific fine-tuning toward dynamic, self-evolving environments. By encoding repository routing knowledge into plain-text catalogs through an adversarial LLM optimization loop, Libra creates a persistent, model-agnostic artifact that continuously improves based on interaction failures. Our findings establish that a system trained entirely on self-generated, synthetic queries can successfully transfer its learned routing capabilities to solve real-world software engineering tasks. Ultimately, Libra demonstrates that optimizing text-based indices via interaction feedback offers a viable and adaptable alternative to traditional retrieval methods.

\subsection{Limitations and future work}
\label{sec:limitations}

While our results are promising, several limitations present opportunities for future work. 
First, the LLM optimization loop is computationally intensive; exploring more efficient Healer designs and catalog structures could reduce this overhead. 
Second, our evaluation assumes static repositories. Because real-world codebases evolve, developing mechanisms for versioning and incremental catalog updates is crucial for practical deployment. 
Third, the current Prompter primarily generates single-hop queries; synthesizing high-quality, multi-hop queries would better reflect real-world complexity and provide a stronger training signal. 
Finally, while we focused on software engineering benchmarks using a specific three-agent architecture, extending Libra to other domains (e.g., general documentation) and exploring the broader design space of agent interactions remain important directions for future research.

\bibliographystyle{plainnat}
\bibliography{references}

\appendix
\section{File vs.\ function accuracy correlation}
\label{app:file-vs-func}

\begin{figure}[h]
  \centering
  \includegraphics[width=0.6\linewidth]{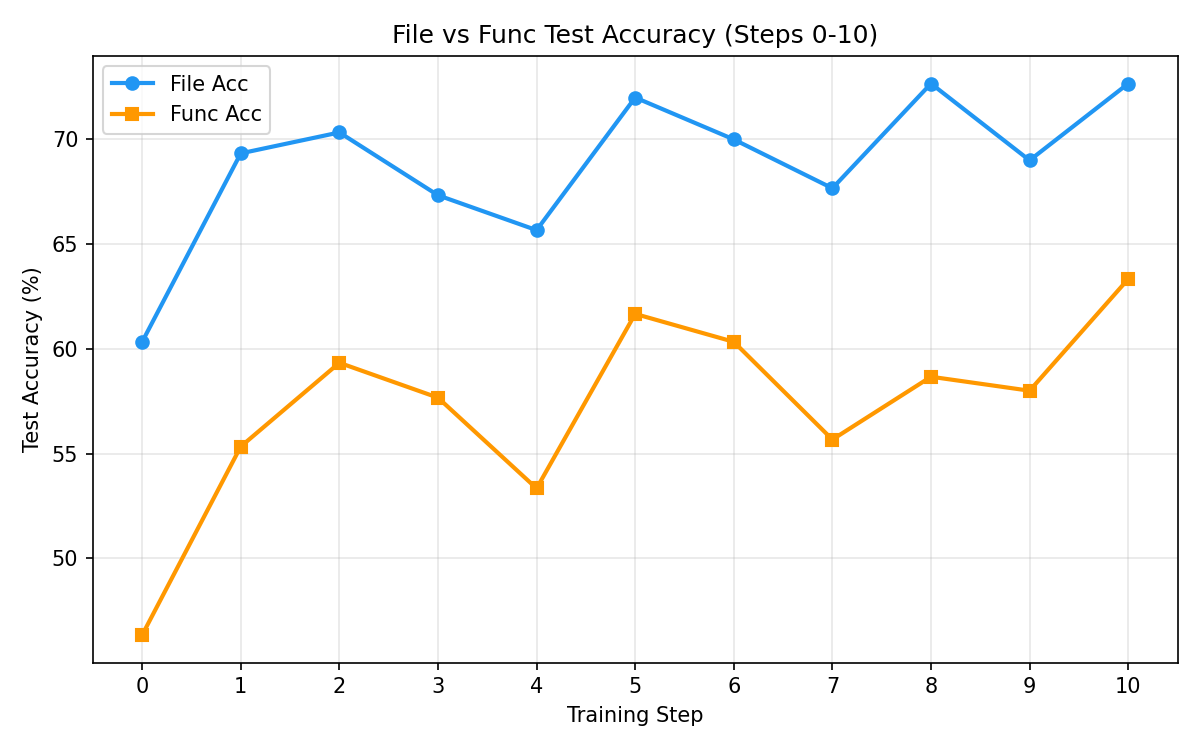}
  \caption{File vs.\ function accuracy on the \texttt{sympy} held-out
  test set across the first $10$ training steps of the trajectory in
  \S\ref{sec:exp:main}. The two metrics are highly correlated
  ($r > 0.95$); we report only file accuracy hereafter.}
  \label{fig:file-vs-func}
\end{figure}

Figure~\ref{fig:file-vs-func} reports test-set file and function accuracy
along the \texttt{sympy} training trajectory of \S\ref{sec:exp:main}
(GPT-5-mini Solver, $\texttt{max\_turns}=5$); the two curves move together
across every step. Note that the \texttt{sympy} train and test splits
were generated by two slightly different Prompter variants
(footnote~\ref{fn:sympy-prompter}), so the file/function gap visible here
is a property of the test-time gold annotations and not of the training
signal. Identical correlation behaviour holds on the other $11$
repositories, where train and test are produced by the same Prompter; we
report file accuracy as our primary metric throughout the paper.

\section{Saturation analysis on recent models}
\label{app:saturation}

We chose GPT-5-mini at $\texttt{max\_turns}=5$ as the training-time
Solver after observing that frontier models saturate the benchmark
even with no catalog. Table~\ref{tab:saturation} reports the
no-catalog baseline for several recent models
on the same $300$-instance test set, all run through the Libra Solver agent at $\texttt{max\_turns}=20$. Above $\sim 85\%$ file-level
accuracy, methodological gains from training become hard to
distinguish from noise, and the portion of failures that are due to the model giving an alternative valid fix becomes significant. We therefore picked a model for which a tightly-budgeted
($5$-turn) configuration still leaves substantial headroom on this
test set. 

\begin{table}[h]
  \centering
  \caption{No-catalog baselines on the Prompter generated $300$-instance test set
  for sympy.}
  \label{tab:saturation}
  \begin{tabular}{lcc}
    \toprule
    Solver               & file@1 & func@1 \\
    \midrule
    Gemini-3-Flash ($20$ turns)       & $86\%$ & $83\%$ \\
    Claude-4.6-Haiku ($20$ turns)     & $88\%$ & $78\%$ \\
    GPT-5-mini ($20$ turns)           & $86\%$ & $82\%$ \\
    \midrule
    GPT-5-mini ($5$ turns, our choice) & $60.3\%$ & $46.3\%$ \\
    \bottomrule
  \end{tabular}
\end{table}

\section{Baseline implementation details}
\label{app:baselines}

This appendix expands on the LocAgent~\citep{locagent2025} and
RepoMem~\citep{repomem2025} baselines summarized in
\S\ref{sec:exp:setup}. Both share Libra's overall agentic-localization
setup (same Solver model, same turn budgets, same gold targets) but
differ in the \emph{retrieval substrate} the Solver is given: LocAgent
exposes a structure-aware repository graph in place of
\texttt{Bash}/\texttt{Read}, while RepoMem layers an additional set of
frozen commit-history tools on top of LocAgent. Crucially, neither
substrate is rewritten during the run, which is the orthogonal axis
Libra adds.

\paragraph{LocAgent.}
LocAgent builds a directed heterogeneous graph over the repository
(nodes: directories, files, classes, functions; edges: contain, import,
invoke, inherit) and layers a sparse, hierarchical entity index on top
of it: (1)~an entity-ID lookup keyed by fully qualified name, (2)~an
entity-name dictionary keyed by short name (with case-insensitive and
\texttt{Class.method} splitting fallbacks), (3)~a BM25 inverted
index~\citep{robertson2009bm25} over entity IDs as a fuzzy fallback
when (1)/(2) miss, and (4)~a BM25 inverted index over each entity's
code chunks for keywords that never appear in any ID (e.g.\ global
variables). The Solver receives three tools in place of Libra's
\texttt{Bash}/\texttt{Read}:
\begin{itemize}
  \item \texttt{SearchEntity} --- keyword search through the
        four-layer cascade above; returns matching entities with their
        source at one of three detail levels (fold / preview / full)
        chosen by match count.
  \item \texttt{TraverseGraph} --- type-aware multi-hop BFS along
        contain/import/invoke/inherit edges.
  \item \texttt{RetrieveEntity} --- full source retrieval for a chosen
        node.
\end{itemize}
We use LocAgent's reference repository defaults: each BM25 index is
queried at \texttt{similarity\_top\_k}=10, with display further capped
at the top 5 hits per query (and at top 3 for non-file/-directory
entities in the entity-ID layer). When the entity-ID BM25 layer
returns nothing, a small \texttt{rapidfuzz} token-set retrieval
(top-3) is consulted before falling through to the content-BM25
layer.

\paragraph{RepoMem.}
RepoMem extends LocAgent with four commit-history tools that expose
\emph{frozen} memory mined offline from the repository's git log:
\begin{itemize}
  \item \texttt{SearchCommit}, \texttt{ExamineCommit} --- episodic
        memory, BM25-matched against commit messages from a window of
        $7{,}000$ prior commits.
  \item \texttt{ViewSummary}, \texttt{SearchSummary} --- semantic
        memory, LLM-generated summaries of the $K{=}200$ most
        frequently edited files.
\end{itemize}
The memory itself is built once before evaluation and never updated by
the Solver; failures cannot feed back into the index. This contrasts
with Libra, where the catalog is the artifact under training and is
continuously rewritten by the Healer in response to Solver failures.

\section{Release artifacts}
\label{app:datasets}

We release code and Prompter data to facilitate reproducibility. The top-level \texttt{README.md}
contains paste-ready commands for every workflow. The code and data are released under the CC BY NC 4.0 License, see README and LICENSE for details.

\paragraph{Code.}
\begin{itemize}
  \item \texttt{agents/} --- the three frozen agents (Prompter,
        Solver, Healer) and their shared \texttt{Read}/\texttt{Bash}
        tools. Each role has an OpenAI-compatible backend
        (\texttt{*\_plain.py}) and a ClaudeAgentSDK backend
        (\texttt{*\_ant.py}).
  \item \texttt{orchestrator/} --- the training loop
        (\texttt{train.py}, implementing Algorithm~\ref{alg:libra})
        and the single-shot replay evaluator (\texttt{evaluate.py}),
        together with their YAML configs (\texttt{train\_config.yaml},
        \texttt{eval\_config.yaml}) which expose every hyperparameter
        used in the experiments.
  \item \texttt{scripts/init\_catalogs.py} (seeds the empty
        \texttt{catalog.md} files the Healer rewrites) and
        \texttt{scripts/replay\_test\_eval.py} (replays the test set
        against any historical catalog snapshot).
  \item \texttt{llm.py} (LLM client), \texttt{pyproject.toml} +
        \texttt{uv.lock} (pinned environment),
        \texttt{.env.example}, \texttt{README.md}.
\end{itemize}

\paragraph{Data ($13$ configs, $\approx\!86$\,MB Parquet).}
\begin{itemize}
  \item \texttt{prompter\_<repo>} ($12$ configs, one per
        \textsc{SWE-bench Lite} repository) --- synthetic Q/A pairs from the Prompter
        at the per-repo base commit listed in Table~\ref{tab:cost}.
        Each config has a \texttt{train} and a \texttt{test} split
        . See Table~\ref{tab:cost} for
        the per-repository instance counts ($90{,}736$ train and
        $3{,}287$ test in total). Schema:
        \texttt{(instance\_id, problem\_statement, gold\_files,
        gold\_functions, gold\_reasoning, chunk\_content,
        line\_numbers, is\_valid\_chunk)}.
  \item \texttt{SWE-bench\_Lite\_Libra} ($1$ config, $199$
        instances) --- the $199$ \textsc{SWE-bench Lite} bug reports that
        re-anchor on a single base commit per repository
        (Table~\ref{tab:cost}), augmented with
        \texttt{gold\_files}/\texttt{gold\_functions} derived from
        the ground-truth patches.
\end{itemize}

\section{Training cost breakdown}
\label{app:cost}

Table~\ref{tab:cost} reports the one-time Prompter data-generation
cost and the per-epoch training cost for each of the 12
\textsc{SWE-bench Lite} repositories. The Prompter and Healer both run Claude Opus~4.6;
the Solver runs GPT-5-mini at 5 turns. We split the training cost
into Train-Eval (Solver evaluation on each training batch) and
Healer (catalog-rewrite proposer + writer agents); their sum is
the total Training cost reported in the rightmost column.
Test-set evaluation cost (the periodic test-eval that runs every
$N$ batches during training, plus any standalone test-eval) is
excluded. Every repository trains for exactly one epoch. All dollar
figures in this section are computed from the per-token list prices
in Table~\ref{tab:prices}.

\begin{table}[h]
  \centering
  \caption{List prices (USD per 1M tokens) used to compute every
  cost figure in this section. \texttt{cache\_r} is the cached-input
  read rate; Anthropic additionally charges a one-time
  cache-creation rate (\$6.25/1M for Opus~4.6) which we fold into
  the input column when relevant.}
  \label{tab:prices}
  \begin{tabular}{lrrr}
    \toprule
    Model              & Input & Output & Cache read \\
    \midrule
    GPT-5-mini         & \$0.25 & \$2.00  & \$0.02 \\
    Claude Opus 4.6    & \$5.00 & \$25.00 & \$0.50 \\
    \bottomrule
  \end{tabular}
\end{table}

\begin{table}[h]
  \centering
  \caption{Per-repository training parameters and cost breakdown.
  \emph{Step size} is the number of instances per healer batch.
  \emph{Base Commit} is the single pinned commit per repo used for
  training and evaluation. \emph{Training} = Train-Eval + Healer.}
  \label{tab:cost}
  \resizebox{\textwidth}{!}{%
  \begin{tabular}{lrrl rrr rrrr}
    \toprule
    Repository & Python LOC & SWE-Lite inst. & Base Commit & Train inst. & Test inst. & Step size & Prompter (\$) & Train-Eval (\$) & Healer (\$) & Training (\$) \\
    \midrule
    sympy               & 431{,}897 & 52 & \texttt{8dcb12a6} & 34{,}000 & 300 & 400 & ${\sim}2{,}373$\textsuperscript{$*$} & 366.59 & 647.68 & 1{,}014.27 \\
    astropy             & 234{,}801 &  3 & \texttt{c76af9ed} &  4{,}670 & 300 & 300 & 335.50 &  27.74 &  98.07 &  125.81 \\
    scikit-learn        & 176{,}926 & 11 & \texttt{c753b77a} &  9{,}589 & 300 & 400 & 649.08 &  52.56 & 100.21 &  152.77 \\
    matplotlib          & 170{,}918 & 15 & \texttt{b7d05919} &  9{,}042 & 300 & 200 & 678.87 &  73.98 & 122.97 &  196.95 \\
    django              & 121{,}189 & 78 & \texttt{e7fd69d0} &  9{,}543 & 300 & 400 & 603.84 &  48.64 & 119.54 &  168.18 \\
    xarray              &  72{,}670 &  5 & \texttt{863e4906} &  4{,}677 & 300 & 300 & 315.49 &  21.50 &  50.07 &   71.57 \\
    sphinx              &  64{,}385 & 10 & \texttt{752d3285} &  4{,}624 & 300 & 200 & 360.39 &  27.30 &  74.87 &  102.17 \\
    pylint              &  39{,}000 &  3 & \texttt{182cc539} &  4{,}601 & 300 & 300 & 356.95 &  25.30 &  59.46 &   84.76 \\
    seaborn             &  28{,}851 &  4 & \texttt{94621cef} &  2{,}401 & 266 & 200 & 206.48 &  10.05 &  24.80 &   34.85 \\
    requests            &  22{,}957 &  2 & \texttt{a0df2cbb} &  1{,}998 & 222 & 150 & 234.31 &   8.85 &  42.02 &   50.87 \\
    pytest              &  21{,}551 & 13 & \texttt{4a2fdce6} &  4{,}694 & 300 & 300 & 328.15 &  22.84 &  31.93 &   54.77 \\
    flask               &   8{,}299 &  3 & \texttt{d8c37f43} &    897   &  99 & 150 &  62.77 &   2.75 &   5.71 &    8.46 \\
    \midrule
    \textbf{Total} & \textbf{1{,}393{,}444} & \textbf{199} & --- & \textbf{90{,}736} & \textbf{3{,}287} & --- & $\mathbf{{\sim}6{,}505}$\textsuperscript{$*$} & \textbf{688.10} & \textbf{1{,}377.33} & \textbf{2{,}065.43} \\
    \bottomrule
  \end{tabular}%
  }
  \\[2pt]
  {\footnotesize\textsuperscript{$*$}Sympy Prompter data generation predates per-call cost instrumentation; the value is estimated by extrapolating from the other repos.}
\end{table}

\begin{figure}[h]
  \centering
  \includegraphics[width=\linewidth]{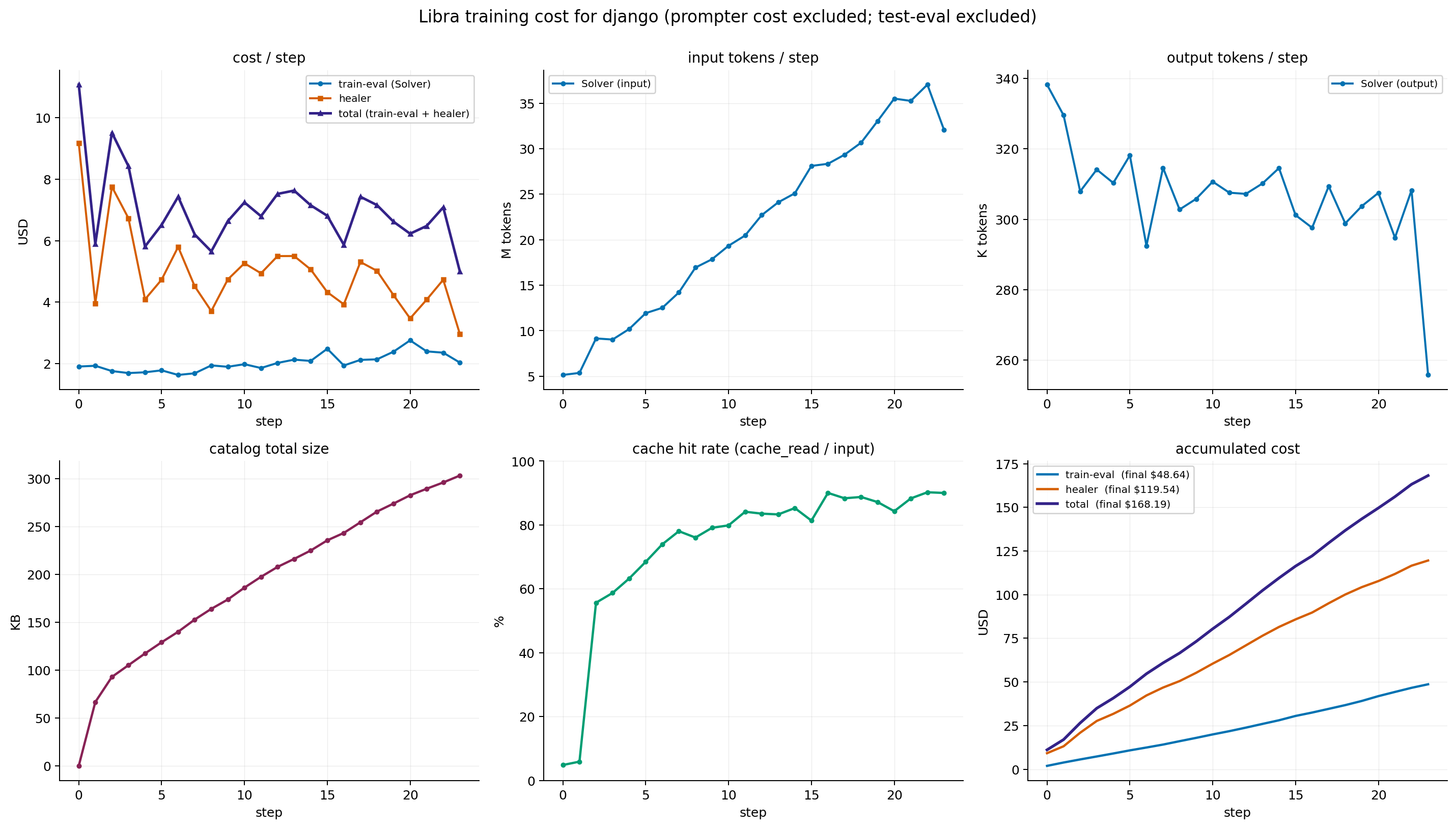}
  \caption{Per-step training cost on django. \emph{Top row}:
  Train-Eval / Healer / total \$ per step (left), Solver input
  tokens (middle), Solver output tokens (right).
  \emph{Bottom row}: cumulative catalog size in KB (left), Solver
  cache-hit rate \texttt{cache\_read / input} (middle),
  accumulated cost (right; final values
  Train-Eval $\$48.64$, Healer $\$119.54$, total $\$168.18$,
  matching the django row of Table~\ref{tab:cost} modulo
  cent-level rounding).}
  \label{fig:django-cost}
\end{figure}

\paragraph{Zoom: per-step cost on django.}
Figure~\ref{fig:django-cost} unpacks the django row of
Table~\ref{tab:cost} step-by-step. The root catalog is always provided in the system prompt, so as it grows, we notice that the Solver's input token grows accordingly. However, the cache-hit rate also climbs as the root catalog stays the same across queries within a batch. As a result, the Train-Eval cost stays relatively flat. Healer cost dominates the bill as we're using the higher-end Claude Opus 4.6 model to rewrite the catalogs.

\section{Training dynamics: additional case studies}
\label{app:cases}

For reference, we walk through two healed-stable test instances
from the top-left cluster of Figure~\ref{fig:case-study}. For
each instance the \emph{trajectory} string records the
evaluation outcome on that instance every $5$ training steps
(at steps $\{0,5,\ldots,85\}$), \texttt{T} for a file-level pass
and \texttt{F} for a fail, with bars separating the early
($0{-}10$), transition ($15{-}35$) and late ($40{-}85$) windows.
The \emph{trigger} and \emph{diff} we list are the
training-batch failure and the resulting \texttt{catalog.md}
edit that we believe fixed the instance stably going forward.

\subsection*{1.\ \texttt{prompter\_173} --- \texttt{\_solve\_system} (nonlinear path vs.\ linear-path trigger)}
\textbf{Test.}
Gold \texttt{solvers/solvers.py::\_solve\_system}; trajectory
\texttt{FFF|FFTFF|TTTTTTTTTT}; question:
\emph{``When iteratively solving a system of equations symbol by
symbol, what happens if a candidate solution for one variable
contains references to variables that were already determined in
earlier iterations?''} (the \emph{nonlinear} branch).

\textbf{Trigger (heal step 20).}
\texttt{prompter\_8387}; same gold function but a \emph{different
sub-path}: \emph{``How does the system of simultaneous first-degree
symbolic equations get converted into an augmented coefficient table
before being dispatched to a linear-system resolution routine?''}
(the \emph{linear} branch). Solver predicted
\texttt{solvers/solveset.py::linear\_eq\_to\_matrix}.

\textbf{Diff (\texttt{solvers/catalog.md}).}
\begin{lstlisting}[basicstyle=\ttfamily\scriptsize,breaklines=true,frame=single]
- `_solve_system(exprs, symbols, **flags)` (L1631-L1829) -
-   Solves systems of equations. For nonlinear polynomial systems
-   with more unknowns than equations, enumerates subsets of free
-   symbols (sized to match equation count), calls
-   `solve_poly_system` on each subset, ...
+ `_solve_system(exprs, symbols, **flags)` (L1631-L1829) -
+   Solves systems of equations.
+   - Linear path: when all poly expressions are first-degree,
+     constructs an augmented coefficient matrix (n x (m+1)) by
+     iterating monomial terms, placing each degree-1 coefficient at
+     the corresponding slot ...
+   - Nonlinear path: for polynomial systems with more unknowns than
+     equations, enumerates subsets of free symbols (sized to match
+     equation count), calls `solve_poly_system` on each subset,
+     and discards candidate solutions that re-introduce already-
+     eliminated symbols.
\end{lstlisting}
The trigger was about the \emph{linear} branch; the test query lands
on the \emph{nonlinear} branch via the same now-richer entry.

\subsection*{2.\ \texttt{prompter\_157} --- \texttt{lib\_interval.cos} via wholesale interval-math enumeration}
\textbf{Test.}
Gold \texttt{plotting/intervalmath/lib\_interval.py::cos};
trajectory \texttt{FFT|TTTTT|TTTTTTTTTT} (a clean flip from no
useful prediction at step 5 to steady-state from step 10);
question:
\emph{``I'm getting incorrect results when evaluating the cosine
of a plain numeric value (not a range) through the interval math
plotting module. The output looks like it's returning the sine
value instead. Where is this likely implemented?''}

\textbf{Trigger (heal step 5).}
\texttt{prompter\_2097} and the near-duplicate
\texttt{prompter\_2279}; both have gold
\texttt{plotting/intervalmath/lib\_interval.py::cosh}
(\emph{different function}, same file): \emph{``When computing the
hyperbolic cosine over a range that crosses zero (e.g., from a
negative to a positive value), how does the interval arithmetic
library determine the lower bound of the result?''} Solver
predicted \texttt{core/expr.py::\_eval\_interval} for the first and
\texttt{functions/elementary/hyperbolic.py::Cosh.\_eval\_interval}
for the second --- both wrong files in the wrong submodule.

\textbf{Diff (\texttt{sympy/catalog.md}, root catalog).}
\begin{lstlisting}[basicstyle=\ttfamily\scriptsize,breaklines=true,frame=single]
  - `intervalmath/` - Interval arithmetic engine (`interval()`)
    for adaptive implicit-plot sampling; used by `plot_implicit`.
+   - `lib_interval.py`: interval-aware math functions (`sin`,
+     `cos`, `cosh` with zero-crossing minimum detection, `exp`,
+     `log`, `atan`, etc.).
\end{lstlisting}
Even though both trigger queries asked specifically about
\texttt{cosh}'s zero-crossing minimum logic, the Healer's edit was
written at the file level: a single bullet enumerating the
interval-aware math functions in \texttt{lib\_interval.py},
including \texttt{cos} verbatim. The test query about
\texttt{cos} sin/cos confusion in the interval-math module then
flips on the very next evaluation step (heal step $5 \to$ test
step $10$) purely because the catalog now mentions
\texttt{lib\_interval.py::cos} by name.

\section{Agent prompts}
\label{app:prompts}

This appendix reproduces verbatim the prompts used for the three
frozen agents in our main experiments. 

\lstdefinestyle{prompt}{%
  basicstyle=\ttfamily\scriptsize,
  breaklines=true,
  frame=single,
  columns=fullflexible,
  keepspaces=true,
  upquote=true,
  showstringspaces=false,
  extendedchars=true,
  inputencoding=utf8,
  literate={…}{{\ldots}}1
           {’}{{'}}1
           {‘}{{'}}1
           {“}{{"}}1
           {”}{{"}}1
}

\subsection{Prompter}
\label{app:prompts:prompter}

The Prompter receives a $\sim\!100$-line code chunk together with a
randomly-sampled additional rule and emits a JSON object containing
a developer-style question, a \texttt{file::function} answer, and a
boolean \texttt{is\_valid\_chunk} flag (used to drop pure-boilerplate
chunks).

\paragraph{System prompt (\texttt{prompter\_system}).}
\begin{lstlisting}[style=prompt]
You generate evaluation Q/A pairs from code chunks. Each pair tests whether an agentic RAG system can locate the correct file and function for a developer's query based purely on semantic understanding, NOT keyword matching.

The chunk you receive is the ground-truth location. Your job: write a question a real developer would ask that this chunk answers, then give the answer as a file::function locator.

### QUESTION STYLES (vary across these)
- "How does the system handle [concept]?" - targets a specific mechanism
- "I got [error/symptom] when [operation]" - bug report / debugging
- "How to support [use case] in [area]?" - feature extension
- "Where is [behavior] implemented?" - localization

### RULES
1. **No self-reference.** Never say "in this snippet", "the provided code", etc. Write as if you're a developer querying a codebase.
2. **Ground the question in the enclosing file.** You may explore beyond the provided chunk - read surrounding code in the same file, or read other parts of the repo for context. However, the final question must be answerable by the file that contains the provided chunk.
3. **Target core logic, not names.** Don't just ask "what does this code do?" - ask about a specific detail, edge case, or mechanism inside it.
4. **NO EXACT IDENTIFIERS (CRITICAL).** You must NOT use the exact names of classes, functions, variables, or files in your question. Instead of asking "How does the Wavefunction class determine limits...", ask "How does the system represent quantum states when determining coordinate boundaries...". Force the evaluating agent to use semantic search.
5. **Dead chunks.** If the chunk is pure boilerplate (only imports, whitespace, closing brackets) with no meaningful logic, set is_valid_chunk to false.
6. Strictly follow the additional rules provided by the user if any.

### ANSWER FORMAT
Use the most specific locator visible in the chunk:
- Top-level function: `filepath::function_name`
- Method on a class: `filepath::ClassName.method_name`
- If multiple functions are relevant, pick the primary one.

### REASONING (think step by step before generating)
1. Explore the chunk's enclosing file and module and write a brief summary of the functions of the chunk, file and module in the repo.
2. Identify the key elements in the chunk: functions, classes, logic branches, comments, edge cases.
3. **List Forbidden Words:** Explicitly list the exact class names, function names, and highly specific variable names found in the chunk. You are banned from using these in the question.
4. Pick the most interesting or non-obvious aspect - a bug, an edge case, a design choice, a specific behavior.
5. Write a question targeting that aspect using conceptual synonyms instead of your forbidden words.

### TOOLS
When using tools (Read, Grep, Bash, etc.) to explore the codebase, always use
**relative paths** (e.g. `core/expr.py`), never absolute paths. The working
directory is already set to the repository root.
\end{lstlisting}

\paragraph{User template (\texttt{prompter\_user\_template}).}
\begin{lstlisting}[style=prompt]
Filepath: $filepath
Lines: $start_line-$end_line

--- CHUNK ---
$chunk_content
--- END CHUNK ---
--- ADDITIONAL RULES ---
$additional_rules
--- END ADDITIONAL RULES ---

Generate an evaluation Q/A pair for this chunk. Output raw JSON only - no markdown fences, no extra text.

{
  "reasoning": "Step-by-step reasoning following the system prompt guidelines.",
  "is_valid_chunk": true,
  "question": "A realistic developer query.",
  "answer": "filepath::Class.method or filepath::function",
}
\end{lstlisting}

\paragraph{Additional rules.}
At each call we sample one rule uniformly at random and substitute it
into \texttt{\$additional\_rules}. The two rules used in our runs are:

\begin{lstlisting}[style=prompt]
# rule_banDomainVocab
**NO MODULE/DOMAIN VOCABULARY (CRITICAL).** Do NOT use words that map directly to module or directory names in the codebase. Banned terms include (but are not limited to): $ban_domain_vocabulary. Any word that would return hits if you `grep`-ed it against the codebase's docstrings or comments is BANNED. Replace concrete library jargon with its abstract formal equivalent.
**Examples:**
Instead, describe the concept abstractly or from a user-behavior perspective.
**Examples:**
- BAD: "How does the geometry module compute triangle area?"
  GOOD: "How does the system compute the area of a three-sided planar figure?"
- BAD: "Where is polynomial GCD implemented?"
  GOOD: "Where is the greatest-common-divisor algorithm for symbolic algebraic expressions implemented?"
- BAD: "How does the pretty printer handle matrix display?"
  GOOD: "How does the human-readable output formatter render two-dimensional tabular data structures?"
\end{lstlisting}

\begin{lstlisting}[style=prompt]
# rule_preferEdgeCase
**PREFER EDGE CASES OVER PRIMARY PURPOSE (CRITICAL).** Your question should target a specific conditional branch, edge case, error-handling path, or special-case logic visible in the chunk - NOT the main/obvious purpose of the code. Focus on if/else branches, try/except blocks, boundary checks, type-specific dispatches, or fallback behaviors.
**Examples:**
- BAD: "How does the matrix constructor work?"
  GOOD: "What happens when the builder receives an entry that is itself a two-dimensional array rather than a scalar?"
- BAD: "How does the parser tokenize input?"
  GOOD: "How does the tokenizer recover when it encounters an unterminated string literal at end-of-file?"
- BAD: "Where is the caching layer implemented?"
  GOOD: "What fallback behavior triggers when the cache store reports a connection failure mid-read?"
\end{lstlisting}

\subsection{Solver}
\label{app:prompts:solver}

The Solver receives only the problem statement as the user message;
the root \texttt{catalog.md} is inlined into the system prompt. We
report results with $\texttt{top\_k}=1$ in the main paper; the
$\texttt{top\_k}=5$ variant differs only in output schema.\footnote{In
the released code, the Solver is named \texttt{Locator} (visible in the
template identifier \texttt{locator\_plain\_system\_template} below). The two names
refer to the same agent.}

\paragraph{System prompt (\texttt{locator\_plain\_system\_template}).}
\begin{lstlisting}[style=prompt]
You are a bug-localization agent. Given a problem statement, explore the
repository to find the single most relevant file and function that would
need to be edited to fix the issue.

The repository contains catalog.md files at the project root and within
each module/sub-package. These catalogs summarize the contents and purpose
of their respective directories. The root catalog.md is provided below. Use it to orient yourself, **then drill into module-level catalog.md files as needed**.

## Root catalog.md
$catalog

## Tools available
  - read: read a file by its path (relative to the repo root). Optionally
specify start_line and end_line to read only a portion of the file.
  - bash: run a shell command. ONLY ls, grep, and find are permitted.

## Output format

When you are confident, respond with ONLY a JSON object (no extra text):
{"file": "<relative path>", "function": "<Class.method or function_name>", "reasoning": "<one sentence>"}
\end{lstlisting}

\subsection{Healer}
\label{app:prompts:healer}

The Healer receives one target catalog file and a batch of
Solver failure reports whose \texttt{target\_md} was assigned to that
file by a deterministic path-prefix match. It diagnoses each failure,
optionally drops it, and applies all surviving edits in-place via
\texttt{Read}/\texttt{Edit}/\texttt{Grep}/\texttt{Bash} tools.

\paragraph{System prompt (\texttt{healer\_system\_template}).}
\begin{lstlisting}[style=prompt]
You are a catalog-healing agent. You will receive a batch of failure reports
from a code-localization system - cases where the Solver predicted the wrong
file for a given **Problem Statement**. Each failure also includes a **Gold
Reasoning** - the correct chain of thought that would have led to the right
answer. Your job is to diagnose each failure, identify those that pinpoint a
gap in the catalog, and improve the catalog in-place to close that gap.

You'll be working under the directory $cwd.

# Catalog Spec (the artifact you are healing)
A catalog file is a navigation aid for the Solver: terse, definitive prose
that lets it pick the right file/module without reading the source.

Structure:
- **Module catalog**: top-level sections group `.py` files by role. Each
  `.py` file (linking to source) is a sub-header inside its section.
- **Root catalog**: same shape, but each **submodule** plays the role of a
  `.py` file - one sub-header per submodule with a summary and link to its
  module catalog.
- Optional **Glossary**, **Architecture Overview**, and **Notes** sections
  at the top.
- Test files are excluded.

Per-entry content (file or submodule sub-header):
1. A one-line summary of what the file/submodule does.
2. Functions/classes with **adaptive verbosity** - match detail to complexity:
   - Trivial/obvious - omit entirely.
   - Simple helpers/utils - list name or declaration only.
   - Moderate complexity - name + short summary.
   - High complexity - bullet list of key functionalities, logic flow,
     components, etc.
   Each listed section (class, function, method, etc.) MUST be annotated
   with its source line range in the form `(L<start>-L<end>)` immediately
   after the name/declaration (e.g. `process_batch (L42-L88)`). Verify ranges by reading the source file - do not guess.
3. (Optional) brief "Caveats" note for surprises or easy-to-misuse behavior.

Fixes must **strictly** follow these constraints:
- Be concise.
- Be definitive, not descriptive (assert what something IS, not how it works).
- Respect adaptive verbosity - don't promote trivial entries to verbose ones.
- Only edit the target catalog file specified in the user message.
- Never create new .md files or delete existing ones.
- Any added or modified content must conform to the Catalog Spec above
  (required structure + per-entry content). Do not invent new structural
  conventions that deviate from the spec.
- Do NOT cross-reference entries - perform Sharpen Entry instead.
- Do NOT include implementation details.
- Do NOT include examples.

Maintain these catalog invariants:
- Each bullet/line <= 250 characters; if longer, split into multiple bullets.
- Each section <= 20 bullet points; if more, merge and purge to <= 15 bullets
  keeping only the most discriminating information.

If a failure can only be fixed by violating the above, drop it. Generalizability beats completeness.

You will be given ONE catalog file to focus on. Follow these steps IN ORDER.
Use tools (Read, Edit, Grep, Bash, etc.) freely.

# Steps
## Step 1 - Gather evidence
For each failure in the batch:
- Study **Gold Chunk** - the code the Solver should have found. Optionally
  read the full **Gold File** for more context.
- Read **Predicted File** - what the Solver chose, to see why it looked
  plausible.
- Compare **Gold Reasoning** to **Prediction Reasoning**. Identify precisely
  where the Solver diverged.

## Step 2 - Triage: is the catalog actually to blame?
Before diagnosing, decide whether the catalog is responsible. Drop the
failure (no edit, count as dropped) if ANY of the following hold - these are
**not** catalog issues:
- **Solver reasoning error** - the catalog context was adequate, but the
  Solver failed to reason about or break down the problem correctly.
- **Ambiguous problem statement** - the question is underspecified or
  misleading; no catalog change could have deterministically routed the
  Solver to the gold target.
- **Intra-file confusion** - the problem hinges on implementation details
  inside the file that aren't tied to its high-level role.
- **Out-of-scope target** - the gold file is a test file or otherwise
  excluded from the catalog spec.
- **Solver gives correct alternative** - the predicted file can fix/answer the problem
  just as well as the gold file; it's essentially an alternative answer.
Skip the rest of the steps if the failure is not catalog-attributable.

## Step 3 - Diagnose root cause
For each remaining (catalog-attributable) failure, pick one or more of:
- **Information Gap** - the catalog entry for the gold target is missing the
  key signal (function/class/role) that would have anchored the Solver.
- **Catalog Noise** - the entry is too verbose/over-detailed; the relevant
  signal is buried. Often means verbosity is set too high for the complexity.
- **Narrow Separation** (common) - the gold entry and a confusable sibling
  entry don't distinguish their functionalities clearly.
- **Structural Issue** - entries are mis-grouped, mis-sectioned, or the
  catalog lacks the sections/glossary/caveats the Solver needed.

## Step 4 - Propose fixes
Choose fixes that target the diagnosed root cause. Examples (not exhaustive):
- **Enrich Entry** - add the missing signal so the gold target is
  distinguishable from commonly confused alternatives.
- **Compact Entry** - drop the entry to a lower verbosity tier (e.g. moderate
  -> simple, or high -> moderate); strip implementation noise.
- **Sharpen Entry** - rewrite definitions of similar entries so each one
  asserts what it uniquely owns.
- **Reorganize** - regroup entries into sections that better reflect role.
- **Meta Change** - add/modify glossary, caveats, or appendix entries when
  the disambiguating signal is cross-cutting.

## Step 5 - Apply fixes
Apply all necessary changes to the target catalog file in-place using your
editing tools. Combine insights from multiple failures - if several failures
point to the same gap, make one coherent fix rather than redundant edits. If
there are >5 fixes required, use the agent tool to parallelize fixes with
subagents.
\end{lstlisting}

\paragraph{User template (\texttt{healer\_user\_template}).}
\begin{lstlisting}[style=prompt]
Target Catalog File: $target_md

Below are $n_failures localization failure(s) where Target Catalog File may be
relevant. Diagnose each one and apply any necessary fixes to the target file.

$failures_block
\end{lstlisting}

\paragraph{Per-failure template (\texttt{failure\_template}).}
The user message embeds one block per failure, formatted as:
\begin{lstlisting}[style=prompt]
--- Failure $failure_idx ---
Instance: $instance_id

Problem Statement:
$problem_statement

Gold File:      $gold_files
Gold Function:  $gold_functions

Gold Chunk (lines $line_numbers):
$chunk_content

Gold Reasoning:
$gold_reasoning

Predicted File:     $pred_file
Predicted Function: $pred_func

Prediction Reasoning:
$reasoning
\end{lstlisting}

\end{document}